\documentclass{phyeauth}
\usepackage{graphicx}
\usepackage{amsmath}
\usepackage{amssymb}

\begin{document}

\begin{frontmatter}

\title{Kondo Regime of a Quantum Dot Molecule: A Finite-$U$ Slave-Boson Approach}
\author[address1,address2]{E. Vernek,\thanksref{thank1}}
\author[address1]{N. Sandler,}
\author[address1]{S. E.~Ulloa,}
\author[address2]{and E.~V.Anda}

\address[address1]{Department of Physics and Astronomy, and Nanoscale
and Quantum Phenomena Institute, Ohio University, Athens, Ohio 45701-2979, USA}

\address[address2]{Departamento de F\'{\i}sica, Pontif\'{\i}cia Universidade
Cat\'olica, Rio de Janeiro-RJ, Brazil}

\thanks[thank1]{
Corresponding author. E-mail: vernek@phy.ohiou.edu}

\begin{abstract}
We study the electronic transport in a  double quantum dot structure connected to leads
in the Kondo regime for both series and parallel arrangements. By applying a finite-$U$
slave boson technique in the mean field approximation we explore the effect of level
degeneracy in the conductance through the system. Our results show that for the {\em
series} connection, as the energy difference of the localized dot levels increases, the
tunneling via the Kondo state is destroyed. For the {\em parallel} configuration, we
find an interesting interplay of state symmetry and conductance. Our results are in
good agrement with those obtained with other methods, and provide additional insights
into the physics of the Kondo state in the double dot system.
\end{abstract}

\end{frontmatter}

Transport properties of small structures such as atoms, molecules and quantum dots have
received a great deal of attention in recent years. Electron coherence at these
nanoscales bring out interference effects that are inherent to the system and are
explorable in experiments~\cite{holleitner}. Confinement of electrons in these
structures results in strong Coulomb effects. The combination of Coulomb interaction
and strong coupling to reservoirs, for example, gives rise to interesting spin-singlet
correlations observed at energies of the order of the Kondo temperature $T_K$,
typically in the sub-kelvin regime. This Kondo effect has been extensively studied
theoretically and experimentally in different quantum dot
systems~\cite{holleitner,Sasaki,jeong}. Rich behavior is anticipated when coherence and
correlations compete in different geometries, and we address some of these issues in
this work.

Although theoretical studies have been presented utilizing different
approaches~\cite{Cornaglia,Dong,Busser,Apel}, the physical richness of these systems
deserves more attention. In this paper we focus our efforts on describing the transport
properties of the double quantum dot system connected to leads either in series or in
parallel while in the Kondo regime. We apply a finite-$U$ slave boson mean field
technique~\cite{Kotliar}, which has been shown to capture the interesting physics in
this regime.  We find that the strong interdot coupling either directly or through the
leads, results in interesting {\em sub}- and {\em super-tunneling} transport regimes,
associated with the characteristic bonding and anti-bonding states of a diatomic
molecule.  It is fascinating that these features are projected into the Kondo regime,
and result in unusual interference effects that should be explorable in experiments
with quantum dots.

\begin{figure}[tbp]
\includegraphics*[width=1.0\linewidth]{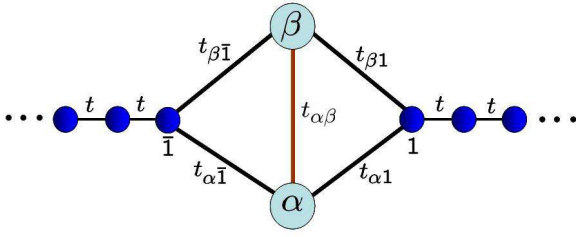}
\caption{Schematic of a double quantum dot coupled to leads.} \label{fig1}
\end{figure}

Figure \ref{fig1} shows a schematic representation of our system.
The transport properties will be studied by calculating the conductance between sites
$-1$ and $1$.  We describe the system by a two-impurity Anderson Hamiltonian
$H=H_{dot}+H_{leads}+H_{dot-leads}$, where $H_{dot}$ describes the physics of the
quantum dots (QDs) before they are coupled to the leads; $H_{leads}$ describes the two
semi-infinite leads, and $H_{dot-leads}$ establishes the couplings. We model the leads
in the tight-binding approximation, which can be described by $H_{leads}=\sum_{\sigma,i
\leq -1}\left(tc^{\dagger}_{i\sigma}c_{i-1\sigma}+H.c.\right) + \sum_{\sigma,i \geq
1}\left(tc^{\dagger}_{i\sigma}c_{i+1\sigma}+H.c. \right)$. In the finite-$U$ slave
boson approach, one enlarges the Hilbert space of the dots by introducing a set of
boson operators, $e_{i}\ (e^\dagger_i)$, $p_{i\sigma}\ (p^\dagger_{i\sigma})$, and $d_i
\ (d^\dagger_i)$, which project onto the empty, singly and doubly occupied electron
states respectively. Within this approach we can write the Hamiltonian $H_{dot}$ as:
\begin{eqnarray}
H_{dot}&=&\sum_{i=\alpha,\beta
\atop\sigma}\left(\epsilon_{i}+eV_g\right)c^{\dagger}_{i\sigma}
c_{i\sigma}+U\sum_{i=\alpha, \beta}d_i^\dagger d_i
\nonumber \\
&&+\sum_{\sigma}\left(t_{\alpha\beta}z^\dagger_{\alpha\sigma}
c^\dagger_{\alpha\sigma}c_{\beta\sigma} z_{\alpha\beta}+H.c \right)\nonumber
\\&&+\sum_{i=\alpha\beta}\left(\lambda^{(1)}_i P_i
+\sum_\sigma\lambda^{(2)}_{i\sigma}Q_{i\sigma}\right),
\end{eqnarray}
where
\begin{eqnarray}
z_{i\sigma}&=&\left(1-d^\dagger_i d_i-p^\dagger_{i\sigma} p_{\sigma}\right)^{-1/2}
\left(e^{\dagger}_ip_{i\sigma}
+p^{\dagger}_{\bar\sigma}d_i\right)\nonumber\\
&&\times\left(1-e^\dagger_i e_i-p^\dagger_{i\bar\sigma} p_{\bar\sigma}\right)^{-1/2}.
\end{eqnarray}
The constraints
\begin{eqnarray}
P_i&=&\sum_\sigma p^\dagger_{i\sigma}p_{i\sigma}+e^\dagger_ie_i
+d^\dagger_id_i-1=0 \\
{\mbox {\rm and}} \nonumber \\
Q_{i\sigma}&=&c^\dagger_{i\sigma}c_{i\sigma}
-p^\dagger_{i\sigma}p_{i\sigma}-d^\dagger_id_i=0
\end{eqnarray}
are enforced in the problem through their respective Lagrange multipliers in $H_{dot}$,
$\lambda^{(1)}_i$ and $\lambda^{(2)}_{i\sigma}$, in order to eliminate unphysical
states. Finally,
\begin{eqnarray}
H_{dot-leads}&=&+\sum_{\sigma, j=\alpha,\beta\atop {j=-1,1}} \left(\tilde
t_{ij}z^\dagger_{i\sigma}c^\dagger_{i\sigma} c_{j\sigma} +H.c.\right)
\end{eqnarray}
allows electrons to travel through the QD region.

In the mean field approximation we replace all the boson operators by their expectation
values. This procedure results in a convenient non-interacting Hamiltonian for the
quasi-electrons with effective energy $\tilde\epsilon_i+\lambda^{(2)}_{i\sigma}$. The
effective Hamiltonian is a function of 14 free parameters which are determined by the
minimization of the free energy $\langle H \rangle$ with respect to all of them. The
information about the strong correlation is however captured by these parameters.
Utilizing the Hellman-Feynman theorem, the condition for minimal free energy requires
${\partial\langle H \rangle}/{\partial x}=0$, where $x$ runs over all the parameters.
This results in a set of non-linear coupled equations that is solved numerically. The
expectation values appearing in these equations are calculated by Green's functions
techniques. Expressions for the Green's functions can be obtained straightforwardly by
the equation of motion method, using the effective mean-field Hamiltonian. From the
Keldysh formalism the equilibrium conductance can be written as $G=4\pi^2t^4\rho_{\bar
1}(\epsilon_F)\rho_{1}(\epsilon_F)|G_{\bar 11}(\epsilon_F)|^2$, where $\rho_{\bar
1}(\omega)=\rho_1(\omega)\equiv\rho_c(\omega)$ is the spectral density of states of the
semi-infinite chain and $G_{\bar 1 1}$ is the propagator that promotes one electron
from site $\bar 1$ to site $1$ and $\epsilon_F$ is the Fermi energy.

\begin{figure}[tbp]
\includegraphics*[width=1.0\linewidth]{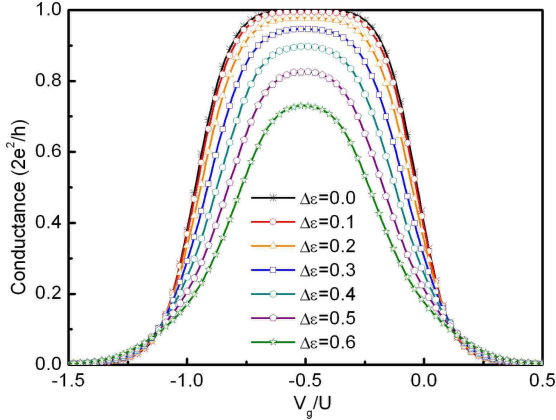}
\caption{(Color online) Conductance as function of gate voltage $V_g$ for various
values of $\Delta\epsilon = \epsilon_\beta - \epsilon_\alpha$, for QDs in series. The
energy unit is $\Gamma$; the parameters are $U=12.5$ and $t_{\alpha\bar 1}=t_{\beta
1}\equiv t^\prime=0.2t$.} \label{fig2}
\end{figure}

One can transform Fig.\ \ref{fig1} into a series configuration if we take $t_{\alpha
1}=t_{\beta\bar1}=0$. The lead bandwidth is $D=4t$ and the Fermi level is set to zero
($\epsilon_F=0$); hereafter we take the broadening $\Gamma$ ($=\pi t^{\prime 2}\rho_c$)
due to the coupling to the leads as the energy unit and show results for temperature
$T=0$. Figure \ref{fig2}
shows the conductance for the case of two dots coupled in series for several values of
the energy difference $\Delta\epsilon=\epsilon_{\beta}-\epsilon_{\alpha}$ and for
$t_{\alpha\beta}=1$. For this relatively small value of $t_{\alpha\beta}$, the role of
the direct interdot connection is only to provide a bridge between the two dots so that
electrons can go through.  Here, the level repulsion between the original dot levels is
essentially negligible (within the level broadening), and the conductance has a maximum
at the particle-hole symmetry condition $(V_g=-U/2)$. For the degenerate case,
$\Delta\epsilon=0$, the dots are in the Kondo regime for the same $V_g$, and the peaks
in the density of states at the Fermi level provide a resonant tunneling condition with
unitary value. This limiting case agrees well with results obtained in
previous works~\cite{Dong}. As $\Delta\epsilon$ increases the dots reach the Kondo regime for
slightly different $V_g$ values and the conductance is suppressed. Larger
$t_{\alpha\beta}$ values result in sizable level repulsion and two conductance peaks.
\begin{figure}[tbp]
\includegraphics*[width=1.0\linewidth]{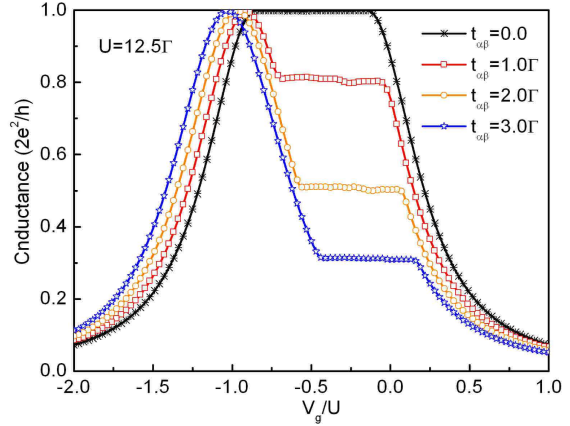}
\caption{(Color online) Conductance as function of gate voltage for various values of
interdot tunneling $t_{\alpha \beta}$. The energy unit is $\Gamma$; the parameters are
$U=12.5$, $\Delta\epsilon=0$ and $t^\prime=0.25t$.} \label{fig3} 
\end{figure}

Figure \ref{fig3} shows the conductance as function of gate voltage for the {\em
parallel} configuration for $\Delta\epsilon=0$ and different values of the interdot
coupling $t_{\alpha\beta}$ Notice that this
configuration resembles the one used in Ref.\cite{Ding} to study the
persistent current of a quantum dot side-coupled to a ring. In this case
however there are two quantum dots embedded in each arm of the ring. For the 
data displayed in Fig. \ref{fig3}, we have chosen  $U=12.5$ and $t_{\beta 1} =
t_{\beta\bar 1}=t_{\alpha 1}=t_{\alpha\bar 1} \equiv t^\prime=0.25t$. For
$t_{\alpha\beta}=0$ we see the typical unitary limit conductance region around
$V_g=-U/2$. In this case we have both dots in the Kondo regime. The half-filled case
results in a total spin $S=1$, provided by the two electrons in the dots. One can
distinguish two well-defined regimes on the conductance curves. The original flat
region around $V_g=-U/2$ for $t_{\alpha\beta}=0$ is split into two regions for larger
interdot coupling.  Let us analyze this behavior in more detail.

\begin{figure}[tbp]
\includegraphics*[width=1.0\linewidth]{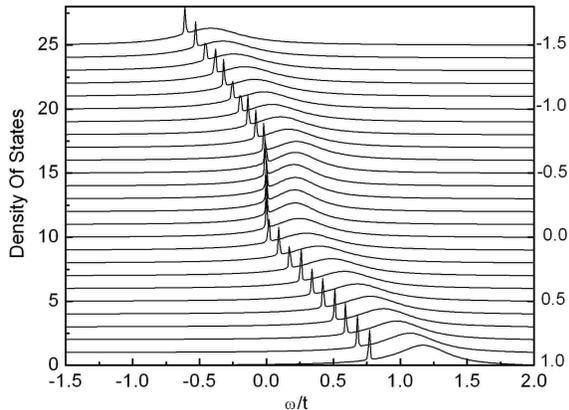}
\caption{Density of states (left axis) as function of energy for a parallel
configuration. Right axis indicate $V_g /U$ values for different curves (offset
vertically). Same parameters as in Fig.\ \ref{fig3} with $t_{\alpha\beta}=2\Gamma$.}
\label{fig4}
\end{figure}

As $t_{\alpha\beta}$ increases, level mixing and repulsion takes place, creating
bonding and antibonding states in the dot system. These coherently-mixed states created
between the two QDs have typical symmetric/antisymmetric configurations, resulting also
in very different effective coupling to the leads. The antibonding state, for example,
contributes a $delta$-like peak to the density of states (see Fig.~\ref{fig4}) and is
completely uncoupled to the leads, due to destructive interference between the two
paths. This interference is akin to the ``subradiant" state discussed in the context of
coupled radiant units, while here is more natural to call it the ``sub-tunneling" state
\cite{Tigran-Alex}. The contribution to the conductance, however, comes from the
bonding state.  The bonding/symmetric state has a much larger effective coupling to the
leads, as in the ``superradiant state", and is therefore termed the ``super-tunneling
state" \cite{Tigran-Alex}. The super-tunneling state results in the broad density of
states feature seen in Fig.\ \ref{fig4} at higher energies.

It is the competition between the bonding and antibonding states
that results in the conductance curves in Fig.\ \ref{fig3}. In the
flat region present for $t_{\alpha\beta}\ne 0$, the antibonding
state is found {\em pinned} around the Fermi level for a large
range of gate voltage, as we can see in Fig.~\ref{fig4}. In that
regime, the contribution to the conductance comes from the tail of
the bonding state that is slightly above the Fermi level. Since
the bonding state is not in resonance with the Fermi level, the
conductance does not reach the unitary limit, until the
antibonding state is filled and moves well below the Fermi level.
Notice that the energy difference between bonding and antibonding
states is determined by $t_{\alpha \beta}$.  As the tail of the
bonding state is increasingly occupied, the sharp antibonding
state reaches the Fermi level.  It remains there until it fills,
while contributing nothing to the conductance.  Notice further
that a change in the {\em sign} of $t_{\alpha \beta}$ produces a
mirror reversal (left to right) of Figs. 3 and 4.  This is in
agreement with the single-particle results in Ref.~\cite{Guevara}.

\begin{figure}[tbp]
\includegraphics*[width=1.0\linewidth]{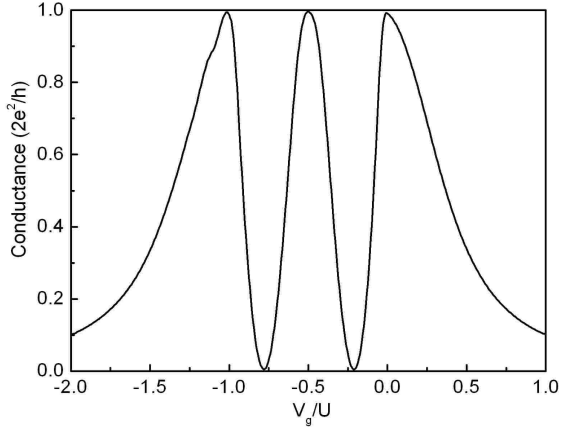}
\caption{Conductance for a parallel configuration for the non-degenerate case. The
parameters are $U=12.5\Gamma$, $\Delta\epsilon=0.075$, $t^\prime=0.1$ and
$t_{\alpha\beta}=0$.}\label{fig5}
\end{figure}

Breaking the degeneracy between dots, $\Delta\epsilon \neq 0$, significantly changes
the conductance of the system.
The conductance for the non-degenerate case is shown in Fig.~\ref{fig5}.  We have
chosen here $t^\prime=0.1t$ and $\Delta\epsilon=0.075\Gamma$, so as to have significant
separation of the levels. We obtain a structure of three clear peaks, analogous to
those obtained in previous works~\cite{Busser,Apel}, where exact diagonalization plus
embedding methods are used to study the Kondo regime. The three peaks can be understood
as follows. The peak around $V_g=0$ is due to electrons travelling through the dot
$\alpha$ with energy $\epsilon_{\alpha}=V_g-\Delta\epsilon/2$, which is in the Kondo
regime ($S=1/2$). As the levels shift with gate, the dot $\beta$ with energy
$\epsilon_{\beta}=V_g+\Delta\epsilon/2$ becomes then accessible and the conductance
starts dropping due to the destructive interference of electrons travelling through the
two different paths, totally vanishing the conductance at $V_g \approx -0.2U$. The peak
at the particle-hole symmetry position ($V_g/U = -0.5$) reaches unity when both dots
are in the Kondo regime ($S=1$). The following valley in the conductance has a similar
interpretation
as the first one. Finally, in the last peak (at $V_g \approx %
-0.8U$) the $\epsilon_{\alpha}$ is well below the Fermi level and the contribution
comes only from the dot $\beta$ in the Kondo regime.

We thank CAPES (Brazil) and NSF-IMC grant 0336431 for support, and Luis Dias da Silva
and Carlos B\"usser for helpful discussions.

\end{document}